**Detecting a large magneto-optical shift in Py/NiO bilayers**

Adrielson Dias[1,2], Maria Clara[1,3], Ozéas Rodrigues[1,3], Gabriel Henrique[1,3], José Lucas [1,2], and José Holanda[1,2,3,4,5,*]

[1]Group of Optoelectronics and Spintronics, Universidade Federal Rural de Pernambuco, 54518-430, Cabo de Santo Agostinho, Pernambuco, Brazil
[2]Programa de Pós-Graduação em Engenharia Física, Universidade Federal Rural de Pernambuco, 54518-430, Cabo de Santo Agostinho, Pernambuco, Brazil
[3]Unidade Acadêmica do Cabo de Santo Agostinho, Universidade Federal Rural de Pernambuco, 54518-430, Cabo de Santo Agostinho, Pernambuco, Brazil
[4]Programa de Pós-Graduação em Física Aplicada, Universidade Federal Rural de Pernambuco, 52171-900, Recife, Pernambuco, Brazil
[5]Programa de Pós Graduação em Tecnologias Energéticas e Nucleares (Proten), Universidade Federal de Pernambuco, Recife, 50740-545, PE, Brazil

## Abstract

Here, magneto-optical magnetometry measurements on Py/NiO bilayers reveal a pronounced magnetic-field-induced wavelength shift, demonstrating strong magneto-optical coupling in this antiferromagnetic system. A systematic and monotonic spectral shift of up to ~ 400 nm is observed as the applied magnetic field increases, saturating at higher fields. Quantitative analysis shows that the associated magneto-optical energy variation is on the order of $10^7$ eV, comparable to the magnon energy scale in NiO. Owing to the large NiO thickness, the observed effect originates from magnons intrinsic to the antiferromagnetic NiO layer rather than from spin currents injected by the Py underlayer. These results provide direct experimental evidence for magnetic-field control of antiferromagnetic magnon energies via magneto-optical interactions, establishing Py/NiO bilayers as a promising platform for optically probing and manipulating antiferromagnetic spin dynamics.

* Corresponding author: joseholanda.silvajunior@ufrpe.br

Orcid iD: https://orcid.org/0000-0002-8823-368X

The recent emergence of antiferromagnetic spintronics has renewed strong interest in antiferromagnetic (AF) materials [1–11]. These materials play a central role in key spintronic devices, most notably spin-valve read heads in hard-disk drives [12]. Traditionally, AF layers have been employed in a largely passive role, serving to pin the magnetization of a reference magnetic layer via interfacial exchange bias [13, 14]. More recently, however, a series of discoveries - including the spin Hall effect in metallic AFs [15–18], the spin Seebeck and spin Nernst effects [19–23], and advances in spin transport in various AF materials [24–30] - have highlighted the active and functional importance of this class of materials. It has become increasingly clear that the distinctive properties of antiferromagnets can be exploited to realize spintronic devices with novel functionalities, such as magnetic memory elements that are highly robust against external magnetic-field perturbations [10, 31, 32] and exhibit ultrafast dynamics compared to ferromagnets [33]. Among antiferromagnetic compounds, NiO is widely regarded as a prototypical room-temperature antiferromagnetic insulator, owing to its simple crystal structure and well-understood spin interactions. The magnetic structure of NiO was established several decades ago [34, 35]. In its paramagnetic phase, NiO adopts the face-centred cubic (fcc) structure characteristic of sodium chloride. Below the Néel temperature $T_N \approx 523$ K, the Ni2+ spins align ferromagnetically within {111} planes along the $\langle 11\bar{2} \rangle$ directions, while adjacent planes are antiferromagnetically coupled through superexchange interactions. NiO has played a pivotal role in experimental studies of exchange bias [12–14, 36, 37], terahertz-frequency magnetic dynamics [38–40], and magneto-optical effects [41–43].

More recently, thin NiO layers incorporated into spintronic heterostructures have been shown to transmit pure spin currents while blocking charge currents [25–30]. In addition, NiO has been demonstrated to generate terahertz-frequency radiation in spin-torque nano-oscillators via magneto-optical interactions [44]. In two-sublattice antiferromagnetic insulators, spin currents carried by the two magnon modes propagate in opposite directions [29, 30]. In uniaxial antiferromagnets such as $Cr_2O_3$, $MnF_2$, or $FeF_2$, these modes are degenerate in the absence of an applied magnetic field, resulting in equal mode populations and a vanishing net spin Seebeck effect [30]. In contrast, NiO crystallizes in an fcc structure with two distinct anisotropy axes: a hard axis along <111> and an easy axis along $<11\bar{2}>$, as illustrated in **Fig. 1(a)**. Below $T_N$, the $Ni^{2+}$ spins are ferromagnetically ordered within {111} planes along $<11\bar{2}>$ directions, with adjacent planes exhibiting opposite magnetization. In this Letter, we investigate the magneto-optical properties of Py/NiO bilayer thin films using magneto-optical magnetometry as the primary experimental technique to probe the interaction between light, magnetic field, and magnetic matter. Surprisingly, we observe a large magneto-optical shift in Py/NiO bilayers, indicating a strong magneto-optical response associated with the AF layer.

The samples were fabricated by sputter deposition onto substrates with lateral dimensions of 6 mm × 2.5 mm, cut from commercial Si wafers (thickness 0.5 mm). Due to natural oxidation, the substrates possess a native $SiO_2$ surface layer approximately 200 nm thick. A 20 nm-thick permalloy (Py) seed layer was first deposited over the entire substrate area by dc magnetron sputtering at ambient temperature. Subsequently, NiO layers with a thickness of 200 nm were deposited by rf magnetron sputtering onto the Py layer. During NiO deposition, an in-plane magnetic field of 300 Oe was applied transverse to the long axis of the strip in order to magnetize the Py layer and promote the growth of well-textured NiO films with a macroscopic AF arrangement. **Fig. 1(b)** shows the X-ray diffraction (XRD) pattern of a 200 nm NiO film deposited on Py, revealing that the film is polycrystalline with a preferred (111) orientation. **Fig. 1(c)** presents a representative energy-dispersive X-ray spectroscopy (EDS) spectrum acquired in conjunction with scanning electron microscopy (SEM) from the Si/Py (20 nm)/NiO (200 nm) bilayer. In **Fig. 1(c)**, the carbon signal originates from the carbon tape used to mount the samples during SEM/EDS measurements, while the silicon peak arises from the substrate. Strong oxygen and nickel peaks confirm the high concentration of these elements in the film. The iron signal is weak and barely detectable, consistent with the limited sensitivity of EDS for elements with concentrations below approximately 10%. Magneto-optical Kerr effect measurements for the Py and NiO/Py samples are shown in **Fig. 1(d)**. The Py layer exhibits a symmetric hysteresis loop centered at zero field, while the NiO/Py bilayer displays a clear shift along the field axis, characteristic of exchange bias and indicative of antiferromagnetic ordering in the NiO layer [24–26]. Together, the structural, compositional, and magneto-optical measurements presented in Fig. 1 confirm that the NiO layers are well textured and exhibit robust antiferromagnetic behavior.

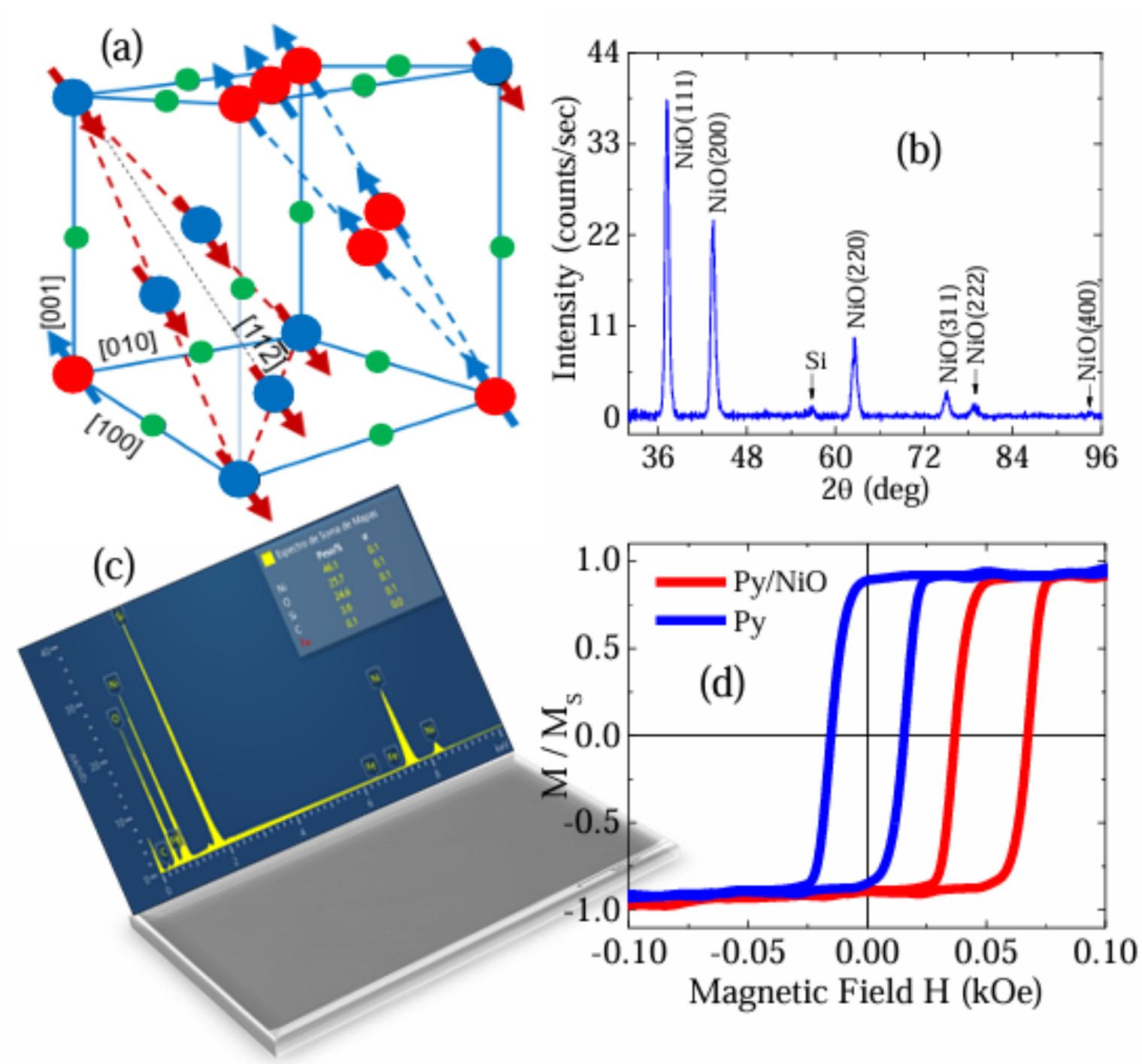

**Figure 1: (a)** NiO unit cell in an fcc structure with two distinct anisotropy axes: a hard axis along <111> and an easy axis along <11$\bar{2}$>; **(b)** shows the X-ray diffraction (XRD) pattern of the NiO film. The diffraction peaks correspond to the crystalline NiO phase, with reflections indexed to the (111), (200), (220), (222), (400), and (311) planes, confirming the formation of polycrystalline NiO with good structural quality; **(c)** energy-dispersive X-ray spectroscopy (EDS) spectrum acquired in conjunction with scanning electron microscopy (SEM); and **(d)** presents the room-temperature magnetic hysteresis loop of a single Py (permalloy) film and Py/NiO bilayer. The loop exhibits typical soft-ferromagnetic behavior, characterized by a narrow hysteresis, low coercive field, and symmetric magnetization reversal. Compared to the Py reference, the magnetic hysteresis loop of the Py/NiO bilayer is shifted along the magnetic field axis, indicating the presence of exchange bias due to interfacial coupling between the ferromagnetic Py layer and the antiferromagnetic NiO layer.

Our magneto-optical magnetometry setup employs a laser with a wavelength of 532 nm and an output power of 5 mW as the light source, a power level chosen to minimize laser-induced heating of the samples. The optical system consists of a sequence of lenses, an iris, a polarizer/analyzer assembly, and a static magnetic-field application setup. The reflected signal is analyzed using a spectrometer coupled to a photodetector (mid-infrared photodiode). A scheme of the experimental setup is shown in **Fig. 2(a)**. Considering the well-textured nature of the NiO films, the corresponding magnon dispersion relations are illustrated schematically in **Fig. 2(b)**. In NiO, the presence of two distinct anisotropy axes lifts the degeneracy of the magnon modes even in the absence of an external magnetic field [32, 33]. This unique characteristic enables NiO to transport spin currents while effectively blocking charge currents over a wide range of magnetic field intensities [32, 34–41]. As emphasized in Ref. [32], the spin current transmitted through NiO decays exponentially with increasing layer thickness, with a characteristic decay length given by the spin diffusion length, which is approximately 7 nm for NiO. Consequently, for a NiO thickness of 200 nm, the spin current due to the optical pump generated in the Py underlayer and reaching the opposite surface of the NiO is reduced by a factor exp(−200/7), rendering it entirely negligible. Under these conditions, the spin current due to the optical pump present at the outer surface of the NiO layer originates predominantly from magnons generated within the well-textured NiO itself. Furthermore, because the magnon dispersion relations in NiO are non-degenerate at zero magnetic field due to its two anisotropy axes [32, 33], a finite magnon energy splitting exists even for $n = k/k_m = 0$ [22]. This results in an energy difference given by $\Delta E_{\alpha\beta} = h\Delta f_{\alpha\beta} = \Delta E_{magnons} = 2.6 \times 10^7$ eV, where $h$ is Planck's constant, $h = 4.3 \times 10^{-15}$ eVs.

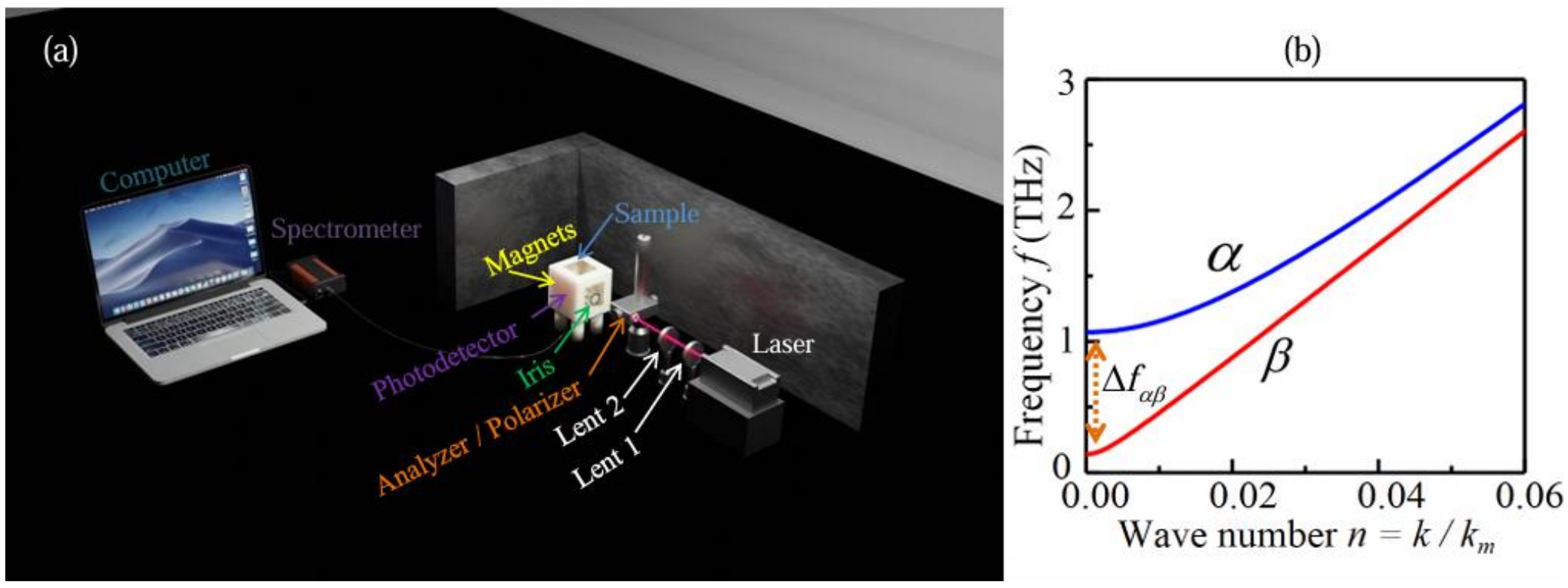

**Figure 2:** Illustrates the experimental setup used for the magneto-optical measurements. Panel **(a)** shows a schematic diagram of the measurement geometry, in which a laser beam is directed onto the sample under an applied magnetic field and the reflected signal is collected by a spectrometer. The schematic highlights the relative positions of the laser source, the sample, and the detection system. Panel **(b)** presents the magnon dispersion relations of antiferromagnetic NiO at T = 300 K near the Brillouin zone center, showing the frequency splitting between the $\alpha$ mode (upper blue curve) and the $\beta$ mode (lower red curve).

Magneto-optical magnetometry measurements were performed on Py (20 nm)/NiO (200 nm) bilayer samples, both in the absence and in the presence of an applied magnetic field. A clear wavelength shift induced by the magnetic field was observed, as shown in **Fig. 3**. This behavior is consistent with previous experimental studies on the band structure of surface magnons in metallic plasmonic crystals composed of a magnetoplasmonic layer, a dielectric spacer, and paramagnetic films [44–50]. In the spectra presented in **Fig. 3 (a)**, the blue curve corresponds to the relative reflectance spectrum measured without an applied magnetic field, while the red curve represents the spectrum obtained under an external magnetic field of H = 5.2 kOe. The application of the magnetic field produces a measurable shift in the spectral features, indicating a pronounced magneto-optical response of the Py/NiO bilayer system. In magnetoplasmonic structures, which are hybrid systems combining noble metals and ferromagnetic materials, the magneto-optical activity is primarily governed by the ferromagnetic component. As a result, significant modifications of the optical response can be achieved with relatively weak applied magnetic fields [47–50]. The analogous behavior observed here suggests that the magnetic-field-induced spectral shift in the Py/NiO bilayers originates from a related magneto-optical mechanism, despite the antiferromagnetic insulating nature of NiO.

Optical systems that exhibit asymmetric propagation between forward and reverse directions are of considerable interest from both fundamental and applied perspectives [32–36]. Such nonreciprocal behavior is a well-established feature of magneto-optical materials and forms the basis of many conventional optical components, including optical isolators. Moreover, structuring magneto-optical dielectrics into magnetic photonic crystals enables enhanced control over the influence of magnetic fields on light propagation. In suitably designed systems, magnetic-field-induced modifications of the photonic band structure can be sufficiently strong to give rise to unidirectional propagation modes. Motivated by these considerations, and by the pronounced magneto-optical response observed in Py/NiO bilayers, we performed systematic magneto-optical magnetometry measurements while varying the applied magnetic field, as shown in **Fig. 3 (b)**. The results presented in **Fig. 3** provide clear and unambiguous evidence for the detection and magnetic-field control of magnon energy via magneto-optical interactions in this system.

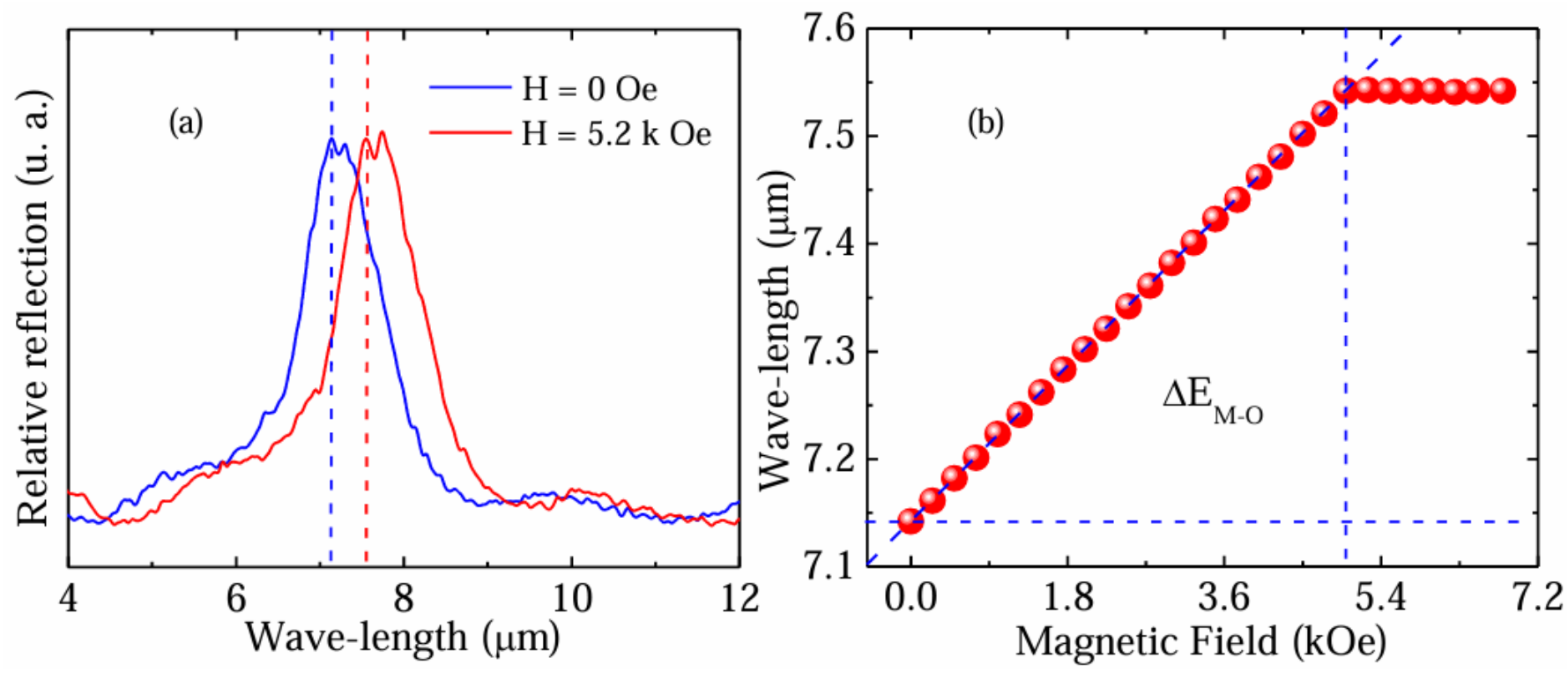

**Figure 3: (a)** (a) Shows the relative optical reflection spectra of the Py (20 nm)/NiO (200 nm) bilayer measured as a function of wavelength under two magnetic field conditions: zero applied field (H = 0 Oe) and an external magnetic field of H = 5.2 kOe. The relative reflectance (in arbitrary units) is plotted over the measured wavelength range. A clear difference between the two spectra is observed, indicating a magnetic-field-dependent modification of the reflected signal and demonstrating that the optical response of the sample is influenced by the applied magnetic field. The laser excitation wavelength used in the experiment was 532 nm. **(b)** Shows the evolution of the spectral position as a function of the applied magnetic field for the Py (20 nm)/NiO (200 nm) bilayer. A clear and systematic shift of the wavelength toward higher values is observed as the magnetic field increases from 0 to 5 kOe and saturates above this value. This monotonic behavior demonstrates a direct correlation between the applied magnetic field and the optical response of the system, ruling out random or thermally induced spectral variations. The magnitude of the wavelength shift reaches approximately $\Delta\lambda \approx 400$ nm, evidencing a pronounced magneto-optical effect.

Our interpretation is based on the quantitative analysis of the magneto-optical energy variation extracted from the data shown in **Fig. 3 (b)**. The magneto-optical energy variation can be expressed as $\Delta E_{M\text{-}O} = \gamma h \Delta\lambda \Delta H / \lambda_{Laser} = 1.1 \times 10^{7}$ eV [41-50], which yields $\Delta E_{M\text{-}O} = 1.6 \times 10^{7}$ eV. In this expression, $\gamma = 2\pi \times 2.8$ MHz/kOe is the gyromagnetic factor [21-30], $\lambda_{Laser} = 532$ nm is the laser wavelength, $\Delta\lambda = 400$ nm is the measured spectral shift, $\Delta H = 5.2$ kOe is the applied magnetic-field variation, and $h = 4.3 \times 10^{-15}$ eVs is Planck's constant. The continuous wavelength displacement observed with increasing magnetic field indicates a magnetically induced modification of the optical response mediated by magneto-optical interactions. This field-dependent spectral shift directly reflects the variation of the magneto-optical energy associated with the excitation and control of magnons in the NiO layer. Given the large NiO thickness (200 nm), the contribution from spin currents injected by the Py underlayer is negligible, confirming that the observed effect originates from magnons intrinsic to the well-textured antiferromagnetic NiO. The data presented in **Fig. 3** therefore provide unambiguous experimental evidence for magnetic-field control of magnon energy through magneto-optical coupling in Py/NiO bilayers. The observed energy shift demonstrates that the magneto-optical energy variation is comparable in magnitude to the magnon energy scale, enabling the optical detection and manipulation of antiferromagnetic magnon modes [39, 50, 52].

In order to elucidate the origin of the magneto-optical interactions associated with the magnon current, we performed Raman spectroscopy measurements using a commercial confocal Raman microscopy system. The setup was coupled to an external magnetic field, and analyzers/polarizers were employed, following the methodology described in Refs. [39] and [50]. Raman spectra acquired in the absence of a magnetic field allow the identification of phonon excitations, corresponding to transverse optical (TO) and longitudinal optical (LO) modes, as well as two-phonon processes, namely 2TO, TO+LO, and 2LO modes. In addition, one-, two-, and four-magnon excitations can also be identified in the spectra [51–55]. **Fig. 4(a)** presents a representative Raman spectrum in which the different excitations are identified. The first four bands are of vibrational origin: the one-phonon TO+LO mode (1F) at 438 $cm^{-1}$, the two-phonon 2TO mode (2F = 2TO) at 631 $cm^{-1}$, the two-phonon TO+LO mode (2F = TO+LO) at 809 $cm^{-1}$, and the two-phonon 2LO mode (2F = 2LO) at 1001 $cm^{-1}$ [51–54]. The most intense band, located at approximately 1400 $cm^{-1}$, is attributed to two-magnon (2M) scattering. The intensity of the 2M mode is intrinsically related to the sample preparation and morphology. Microscopically, the 2M excitation originates from the superexchange interaction between nearest-neighbor $Ni^{2+}$ ions along linear atomic chains in the NiO lattice [51–53]. **Fig. 4(b)** compares Raman spectra measured without an applied magnetic field (blue curve) and under an external magnetic field of H = 5.2 kOe (red curve). The application of the magnetic field induces a clear shift in the Raman spectrum, revealing a pronounced magneto-optical response of the Py/NiO bilayer system. The observed shift has a wavenumber magnitude of approximately 74 $cm^{-1}$, corresponding to a wavelength variation of about 400 nm ($\Delta\lambda \approx 400$ nm), in good agreement with the results presented in **Fig. 3**. Notably, only the two-magnon (2M) mode exhibits a measurable shift, demonstrating that the effect is purely of magnetonic origin and that its detection arises from magneto-optical interactions [49].

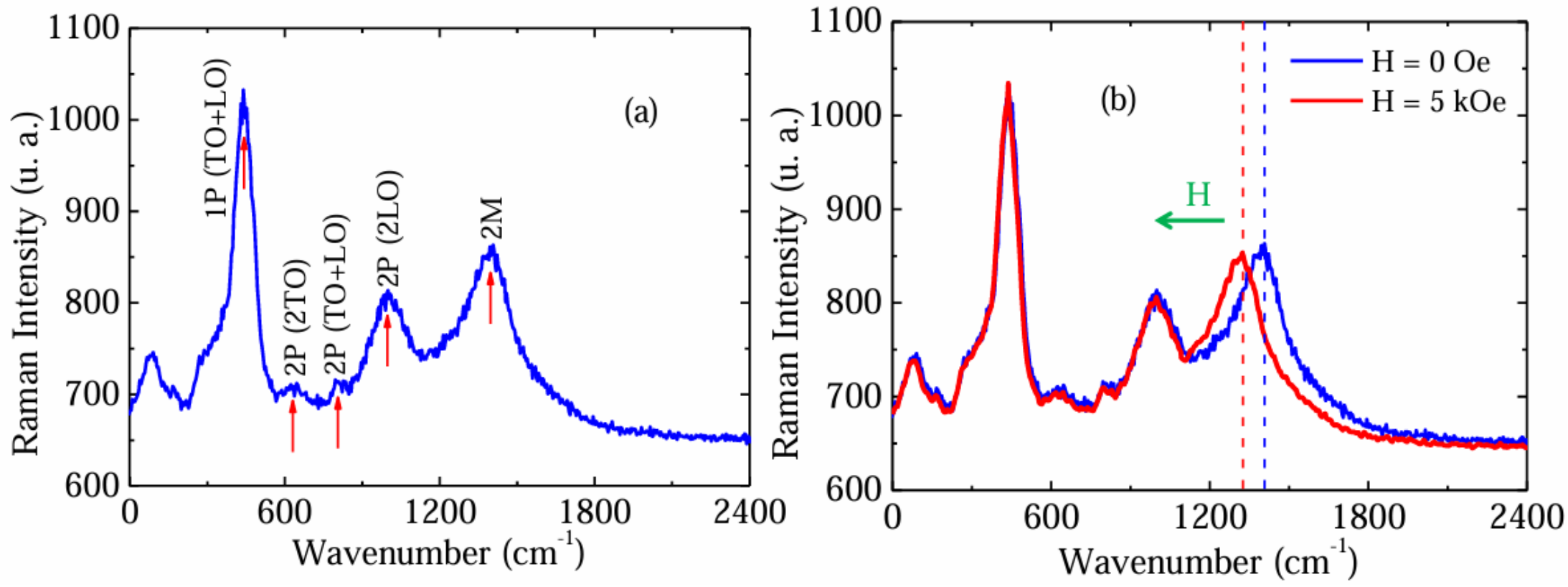

**Figure 4: (a)** Presents the vibrational and magnon Raman peaks. The observed features are assigned to one- and two-phonon scattering processes, including the first-order TO+LO mode and second-order modes such as 2TO, TO+LO, and 2LO, as well as a two-magnon (2M) contribution. **(b)** Shows that the vibrational modes remain unchanged upon application of the magnetic field, while the two-magnon (2M) mode exhibits a shift of approximately 74 cm$^{-1}$ (≈400 nm) under an applied magnetic field of 5 kOe. No additional shift is observed for magnetic fields above this value, indicating saturation of the magneto-optical response.

In summary, we observe a pronounced magneto-optical shift in Py/NiO bilayers originating from a magnon current in the NiO layer. This effect arises from non-degenerate magnon scattering relations that exist even in the absence of an external magnetic field, a direct consequence of the two magnetic anisotropy axes in NiO. As a result, a finite magnon energy is present at zero field, enabling efficient magneto-optical coupling. The systematic field-induced spectral shifts observed in **Figs. 3** and **4** therefore provide unequivocal evidence of strong magneto-optical coupling in the Py/NiO system, with the magnitude of the wavelength shift serving as a quantitative measure of the interaction strength. These findings place the present results within the broader context of intense current interest in antiferromagnetic magnonics and magneto-optical phenomena, underscoring their significance and potential impact.

**Acknowledgements**

This research was supported by Conselho Nacional de Desenvolvimento Científico e Tecnológico (CNPq) with Grant Number: 300631/2025-1, Coordenação de Aperfeiçoamento de Pessoal de Nível Superior (CAPES) with Grant Number: PROAP2025UFRPE, and Fundação de Amparo à Ciência e Tecnologia do Estado de Pernambuco (FACEPE) with Grant Number: APQ-1397-3.04/24.

**Contributions**

A. D., M. C., O. R., and G. H. analyzed all the experimental measures and J. H. discussed, wrote and supervised the work.

**Conflict of interest**

The authors declare that they have no conflict of interest.

**Data Availability Statement**

Data will be made available on reasonable request.